\documentclass[twocolumn,times]{aastex63} 

\usepackage{natbib}
\usepackage{amsmath}
\usepackage{amssymb}
\usepackage{subfigure}
\usepackage{url}
\usepackage{CJK}
\usepackage{longtable}

\renewcommand{\vec}[1]{ {\mathbf #1} }

\newcommand{\Eq}{{Equation}}
\newcommand{\Eqs}{{Equations}}
\newcommand{\Fig}{{Figure}}

\newcommand{\SDO}{{\it SDO}}

\submitjournal{ApJL}
\shorttitle{Force and Torque}
\shortauthors{Duan et al.}

\begin{document}

\title{On the Lorentz Force and Torque of Solar Photospheric Emerging Magnetic Fields}
\correspondingauthor{}
\email{duanaiy@mail.sysu.edu.cn; chaowei@hit.edu.cn}

\author[0000-0002-7018-6862]{Aiying Duan}
\affiliation{Planetary Environmental and Astrobiological Research Laboratory (PEARL), School of Atmospheric Sciences, Sun Yat-sen University, Zhuhai 519000, China}

\author{Chaowei Jiang} 
\affiliation{Institute of Space Science and Applied Technology, Harbin Institute of Technology, Shenzhen 518055, China}

\author{Shin Toriumi}
\affiliation{Institute of Space and Astronautical Science (ISAS)/Japan Aerospace Exploration Agency (JAXA), 3-1-1 Yoshinodai, Chuo-ku, Sagamihara, Kanagawa 252-5210, Japan}

\author{Petros Syntelis}
\affiliation{School of Mathematics and Statistics, St. Andrews University, St. Andrews, KY16 9SS, UK}

\begin{abstract}
Magnetic flux generated and intensified by the solar dynamo emerges into the solar atmosphere, forming active regions (ARs) including sunspots. Existing theories of flux emergence suggest that the magnetic flux can rise buoyantly through the convection zone but is trapped at the photosphere, while its further rising into the atmosphere resorts to
the Parker buoyancy instability. To trigger such an instability, the Lorentz force in the photosphere needs to be as large as the gas pressure gradient to hold up an extra amount of mass against gravity.
This naturally results in a strongly non-force-free photosphere, which is indeed shown in typical idealized numerical simulations of flux tube buoyancy from below the photosphere into the corona.
Here we conduct a statistical study of the extents of normalized Lorentz forces and torques in the emerging photospheric magnetic field with a substantially large sample of \SDO/HMI vector magnetograms. We found that the photospheric field has a rather small Lorentz force and torque on average, and thus is very close to a force-free state, which is not consistent with theories as well as idealized simulations of flux emergence. Furthermore, the small extents of forces and torques seem not to be influenced by the emerging AR's size, the emergence rate, or the non-potentiality of the field. This result puts an important constraint on future development of theories and simulations of flux emergence.
\end{abstract}


\section{Introduction}
\label{sec:intro}
Sunspots and solar active regions (ARs) are believed to be generated by
magnetic flux emerging through the solar surface, i.e., the
photosphere~\citep{Parker1955}. In the first place, magnetic field is generated by
the solar dynamo action in the deep convection zone. Then the
magnetic fluxes are further intensified and adopt the form of flux
bundles because of stretching and twisting of their magnetic field lines
by turbulent fluid motions in the convection zone. Eventually, the
flux bundles are pushed up by magnetic buoyancy and
rise into the photosphere to form ARs.

Since we are not able to observe directly the process in the
sub-photospheric layers, most of the knowledge on flux emergence comes
from theoretical models and numerical MHD simulations~\citep{Archontis2008, Cheung2014, archontis_emergence_2019}. Many
simulations show that, a flux tube, generally twisted, experiences a
typical two-stage emergence process~\citep[e.g.,][]{Toriumi2010} during its passage from the
convection zone into the atmosphere, i.e., the solar corona.  In the
first stage, the magnetic buoyancy force pushes upward the flux tube
through the convection zone~\citep{Parker1955}, but it is trapped at the shallow layers
near the solar surface because of the strongly sub-adiabatic
stratification (i.e., much smaller temperature gradient than an
adiabatic stratification) of the photosphere~\citep[e.g.,][]{Syntelis2019S}. Then in the second
stage, with more and more magnetic flux pileup just below the
surface, the magnetic pressure gradient increases continuously and eventually
the Parker instability~\citep{shibata_nonlinear_1989, archontis_emergence_2004}, a kind of
magnetic Rayleigh-Taylor instability, is triggered to let part of the flux break
through the dense photosphere, which is typically observed as a
continuous magnetic flux increasing on the photosphere during the
emerging of an AR.

To trigger the Parker instability at the sub-adiabatic photosphere,
the Lorentz force must be built up such that it can support against gravity an extra amount of mass comparable to what is supported by the photospheric gas pressure gradient~\citep{newcomb_convective_1961, Acheson1979, archontis_emergence_2004}. Meanwhile the plasma $\beta$
(ratio of plasma pressure to magnetic pressure) is on the order of
unity in the photosphere~\citep{Gary2001, cho_determination_2017}.
Thus, {\it in the photosphere}, the magnitude of Lorentz force should be comparable
to that of the magnetic pressure gradient, and so the emerging field in the photosphere is
far from a force-free state in which the Lorentz force vanishes as a result of the balancing between magnetic pressure force and magnetic tension~\citep{Wiegelmann2012solar}.
Furthermore, once the buoyancy instability is triggered, the emergence runs into a dynamic phase in which there is a strong interaction of the magnetic field with the plasma, and the field cannot be force-free.
However, when the field
emerges into the atmosphere by a few scale heights above the photosphere, it relaxes quickly to a closely force-free state because of the fast decreasing of the plasma density and pressure with height~\citep{Fan2009}. Thus the distribution of the Lorentz force above the solar surface should be restricted mainly within a small height. To quantify the global extent of this force, one can make use of the Gauss's law by expressing the net volumetric Lorentz force above the photosphere in a surface integral of Maxwell stress tensor of the photospheric magnetic field~\citep[][see the formula in Section~\ref{sec:data}]{Molodenskii1969, aly_properties_1984, Low1985, Fisher2012}, which are observable and thus the results can be used to compare with and constrain theories and simulations.
There have been several works done to estimate the global Lorentz
force of ARs using the observed vector magnetograms from different
instruments~\citep{Metcalf1995, moon_forcefreeness_2002, tiwari_force-free_2012, liu_statistical_2013, liu_research_2015}, and some of them conclude that the photospheric
field is actually not far away from force-free~\citep[e.g.,][]{moon_forcefreeness_2002, tiwari_force-free_2012}. However all the previous studies are carried out using the snapshots of developed
ARs without considering the case of flux-emerging ARs.

In this Letter, we perform a systematic survey of the global force focusing on ARs in their full emergence phase and using the vector magnetograms from Solar Dynamics Observatory/Helioseismic and Magnetic Imager (\SDO/HMI)~\citep{Hoeksema2014}, which are both done for the
first time.  Furthermore, in addition to checking the global force, we
also estimate the global Lorentz torque, which has not been considered in
previous investigations. Our results show that the emerging photospheric fields are actually very close to the force-free state, which is not consistent with the
results from typical idealized simulations of twisted flux tube emerging from
below the photosphere. This result puts an important constraint on
the future development of theories and simulations of flux emergence.

\section{Method and Data}
\label{sec:data}

Since the Lorentz force can be expressed as the divergence of a
tensor, the Maxwell stress, the integration of force in a volume can
be expressed as surface integral of the tensor. Furthermore, by
assuming that the emerging magnetic field above the photosphere (i.e., $z>0$) is well isolated and its strength falls off fast enough as going upward to infinity,
the net Lorentz force $\vec F = \int \vec J\times \vec B dV$ in the volume of $z>0$ can be expressed
as integration of the Maxwell stress tensor on the photosphere $z=0$, which is
given by~\citep{aly_properties_1984, Low1985, Fisher2012}
\begin{equation}\label{force}
\begin{split}
  F_x = -\frac{1}{4\pi}\int B_x B_z dxdy,  \\
  F_y =-\frac{1}{4\pi}\int B_y B_z dxdy,  \\
  F_z =-\frac{1}{8\pi}\int (B_z^2-B_x^2-B_y^2) dxdy,
  \end{split}
\end{equation}
We note that the same formula can also apply to the interior volume below the surface (by removing the minus sign before the integrations) if the magnetic field below the surface is also well isolated, which, however, is not generally fulfilled unless the surface integration is taken for the full sphere.

To compare the forces in different magnetic fields with different flux contents, it is better to
use a normalized measurement. We follow~\citet{Low1985} and
\citet{Metcalf1995} by employing the integrated magnetic pressure
force $F_p$ given by
\begin{equation}
  F_p = \left |\int \nabla \left(\frac{B^2}{8\pi}\right) dV \right | = \frac{1}{8\pi}\int (B_x^2+B_y^2+B_z^2) dxdy,
\end{equation}
and the normalized forces are ratios defined as
$f_x = |{F_x}|/{F_p}$, $f_y = |{F_y}|/{F_p}$, and
$f_z = |{F_z}|/{F_p}$, respectively. For a field being close to
force-free, it must have all the ratios much less than unity, i.e.,
$f_x \ll 1$, $f_y \ll 1$, and $f_z \ll 1$~\citep{Low1985}.
\citet{Metcalf1995} suggested that the magnetic field can be
considered as force-free completely if the normalized forces are all less
than or equal to $0.1$, and this criterion is widely accepted by the later studies~\citep{ moon_forcefreeness_2002, tiwari_force-free_2012, liu_statistical_2013, liu_research_2015, jiang_influence_2019}. On the other hand, for a strongly non force-free field, the ratios can be close to unity, meaning that the magnetic pressure force and the tension force are so unbalanced that
the net Lorentz force is comparable to one of its components, the total magnetic pressure
force.
The HMI team~\citep{SunX2019} released a data set, \texttt{cgem.Lorentz}, which also contains the integrated Lorentz forces of the HMI vector magnetograms. A minor difference exists between the horizontal normalized forces we defined here and the ones in \texttt{cgem.Lorenzt}, i.e., $\epsilon_x$ and $\epsilon_y$, and $\epsilon_z$ in \Eq~(4) in \citep{SunX2019}, which are related by $f_x = 2|\epsilon_x|$,  $ f_y=2|\epsilon_y|$, and  $f_z = |\epsilon_z|$. We use the \texttt{cgem.Lorentz} data to verify our calculations.

It should be noted that these ratios being less than $0.1$ is only a
necessary condition for the fields to be force-free; even if the
global integration of force is zero, the forces locally in different
parts are not necessarily zero \textbf{(for example, near the edge of a sunspot, where the local Lorentz force is balanced by gas pressure gradients)}. To put an additional constraint, we can
compute the net torque induced by the Lorentz force, $\vec T = \int \vec r \times (\vec J\times \vec B) dV$ , which is given
as~\citep{Aly1989}
\begin{equation}\label{torque}
  \begin{split}
    T_x = -\frac{1}{8\pi} \int y(B_z^2-B_x^2-B_y^2) dxdy, \\
    T_y = \frac{1}{8\pi} \int x(B_z^2-B_x^2-B_y^2) dxdy, \\
    T_z = \frac{1}{4\pi} \int (yB_xB_z-xB_yB_z) dxdy,
  \end{split}
\end{equation}
where the origin of the coordinates $(x,y)$ is set at the lower left
corner of the magnetogram. Similarly, the normalized
torques are defined as $t_x = |{T_x}|/{T_p}$, $t_y = |{T_y}|/{T_p}$,
and $t_z = |{T_z}|/{T_p}$, and $T_p$ is the magnitude of the net torque induced by only the magnetic
pressure force, $T_p = |\int  \vec r \times (\nabla \frac{B^2}{8\pi}) dV| = \sqrt{T_{px}^2+T_{py}^2}$ where
\begin{equation}
\begin{split}
T_{px} = \frac{1}{8\pi} \int y(B_x^2+B_y^2+B_z^2) dxdy,\\
T_{py} = \frac{1}{8\pi} \int x(B_x^2+B_y^2+B_z^2) dxdy.
  \end{split}
\end{equation}
A very important requirement in using the formula is that the magnetic field has a high-degree
balance of the positive and negative fluxes, which can be quantified
by the ratio of the net flux $\Phi_{\rm n}$ to the total unsigned flux $\Phi_{\rm u}$,
\begin{equation}\label{eflux}
    e_{\rm flux} = \frac{\Phi_{\rm n}}{\Phi_{\rm u}} = \frac{|\int B_z dxdy|}{\int |B_z| dxdy}.
\end{equation}
Last we note that since the formula of net force and torque, \Eqs~(\ref{force}) and (\ref{torque}), are
integrations of the pixels of magnetogram, they are not sensitive to the resolutions of the magnetogram~\citep[see also][]{jiang_influence_2019}; for instance, we found that the results are almost unchanged by reducing resolution of the original magnetogram with a factor of 2 or 4. Furthermore, we found that calculations of the force and torque are not very sensitive to the noise in the vector magnetograms. \citet{Bobra2014} state that, in the HMI vector magnetograms, the field strengths below 220~G are generally considered to be noise. We have also calculated the forces and torques by using only the field above the noise threshold of $220$~G, and found that they are changed very little, likely due to that the contributions of the small-scaled, unresolved fields cancel each other in the integrations.

\begin{figure*}[htbp]
  \centering
  \includegraphics[width=0.6\textwidth]{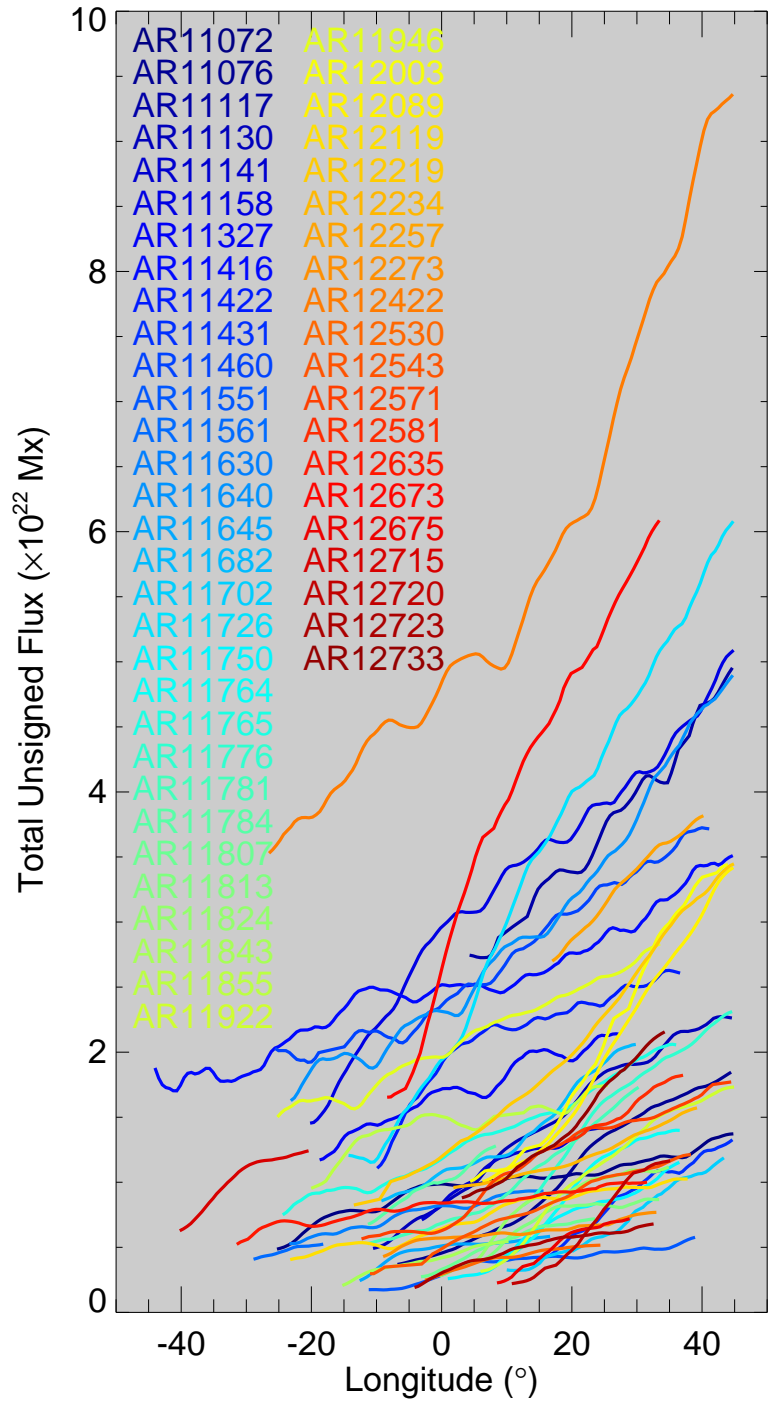}
  \caption{Magnetic flux evolution for all the 51 flux emerging
    ARs. The horizontal axis shows the longitude of AR's center
    with respect to the solar disk center, thus it also indicates the
    evolution time. Note that the events are selected with longitude
    between $-45^{\circ}$ and $45^{\circ}$. Different events are
    plotted in different colors, which denote the AR's number.}
  \label{flux_evolution}
\end{figure*}

We survey all the space-weather HMI AR patches~\citep[SHARP,][]{Bobra2014} with definite NOAA number observed from 2010 May, when {\SDO} began to operate, until the end of 2019. The
following criteria are used to select the ARs. First, we focus on the
emerging phase of ARs from almost nothing on the solar surface to its
peak flux. Thus the ARs should have significant flux emergence during
their passing on the solar disk, i.e., the total unsigned magnetic
flux shows a evolution trend of overall monotone increase. Second, the
target AR should be well separated from the surrounding ARs (presumably
a single flux tube emergence and isolated from interaction
with significant pre-existing field), so the positive and negative
fluxes of the region can be balanced each other in a good degree, and
in particular we only select the ARs with the flux-balance parameter
$e_{\rm flux} < 0.1$ during their emergence phase. Third, to reduce
the observation errors, we only select the duration when the ARs are
located within $\pm 45^{\circ}$ in longitude from the Sun's central
meridian (as will be shown in the next section, the distance of the observed
AR to the solar disk center has a systematic influence on the
magnitudes of the forces and torques).  By all these criteria, we
finally obtained 51 ARs with significant flux emergence during their passage on the
solar disk. The ARs and their observed time durations are listed in Table~\ref{tab1},
including their start, end times and their Carrington coordinates. The
evolutions of total unsigned magnetic fluxes of all the ARs,
calculated based on one-hour cadence SHARP data and smoothed by a
six-hour window, are shown in \Fig~\ref{flux_evolution}. As can be
seen, overall the larger the AR's total flux is, the faster the AR
emerges~\citep{otsuji_statistical_2011, norton_magnetic_2017}.
With all the ARs considered, there are in total
$3536$ vector magnetograms obtained for our study to compute the forces and torques of the photospheric field.

We also take two typical, independently-developed flux-emergence MHD simulations as a comparison with the observations. The two MHD simulations are obtained from \citet{toriumi_numerical_2017} and \citet{syntelis_recurrent_2017},
respectively, for the aim of reproducing the birth of ARs with significant
non-potentiality. Both simulations used typical settings of a twisted
flux tube that is initially placed in the convection zone (with several Mm below the photosphere) and buoyantly rises to the photosphere, where it partially emerges into the corona.
To compare the simulations with observations, it is crucial to use the same physical height of photosphere. We note that in the simulations, the photosphere is slightly lifted by the emerging flux, and the buoyancy instability essentially happens at altitudes of approximately two scale heights~\citep[e.g.,][]{Fan2009, syntelis_recurrent_2017}. Thus, we perform the force and torque computations using the magnetic field at three different heights, starting from $z_0$, which is closest to $z=0$ in each numerical model, and moving upwards to the Parker instability height ($z_2$) at increments of around one pressure scale height.

\section{Results}
\label{sec:res}
 Before giving the \textbf{statistical results}, we first show the evolutions of two well-studied
ARs, NOAA 11158 and 12673, and compare them with the two flux-emergence MHD
simulations. Both the ARs have a fast flux-emergence phase during
their passage on the solar disk and have strong non-potentiality
manifested by shearing and rotating motions of the emerging sunspots~\citep{Sun2012, YangS2017, YanX2018}. As shown in \Fig~\ref{compare}, both the ARs have significant flux increase, with an average
emerging rate of $\sim 5\times 10^{20}$~Mx~h$^{-1}$, and their fluxes
are \textbf{very well} balanced (with $e_{\rm flux} \sim 0.05$). The normalized forces and torques are mostly
less than $0.1$ \textbf{for the full duration}, suggesting that the field is very close to
force-free in the emergence process. Note that for the normalized forces, our results are almost identical to those directly from the \texttt{cgem.Lorentz} data set.

In the simulations, although their flux contents are smaller by
at least two orders of magnitude than the two ARs, their peak emerging rates are comparable the observed ones. Since the two simulations use similar settings, they show very similar evolution, with a very fast rising and saturation of the fluxes in roughly an hour, while the observed ones have a steady flux increase for several days. The very distinct difference between the simulations and the observed ARs lies in the forces and torques: once the photospheric magnetic fluxes begin to increase in the simulations, the forces and torques in all the three levels (see $z_0$, $z_1$ and $z_2$ labeled in the \Fig~\ref{compare}) rise to significantly larger values ($\sim 0.8$ for $f_z$) and torques ($\sim 0.6$ for $t_x$ and $t_y$), attaining the order of unity. This means that in the simulated cases, the balance between the tension and magnetic pressure gradient is strongly destroyed, i.e., in the extremely non-force-free state. For the $z_0$ level, this strongly-forced state occurs during the entire emerging phase, while for a little higher, the force and torque decrease systematically with time. In particular, in \citet{syntelis_recurrent_2017}'s simulation, they decrease rather fast at $z_2$ where the Parker instability happens, reaching almost zero when the flux saturates.
Therefore, our results show that during the flux injection phase, i.e. around the emergence rate peak, the forces and torques in typical flux emergence models are much larger than the observed ones. However later, the dynamics settle to more realistic values.
Note that it is due to the perfect symmetry of two emerging polarities in the simulations that the net forces in horizontal directions (i.e., $f_x$ and $f_y$) and the net torques in vertical direction ($t_z$) are both zero.

\begin{figure*}[htbp]
  \centering
  \includegraphics[width=0.9\textwidth]{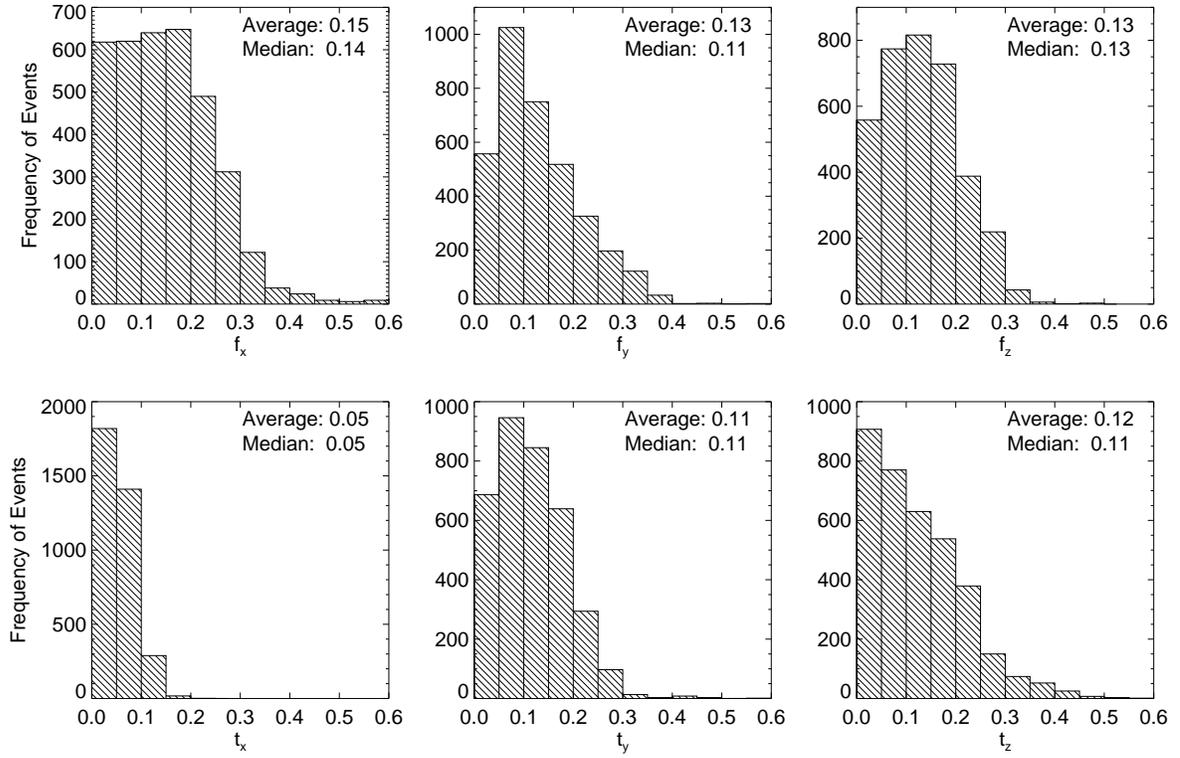}
  \caption{Histogram distributions of normalized Lorentz force
    $f_{x}$, $f_{y}$, and $f_{z}$ (top panels) as well as torque
    $t_{x}$, $t_{y}$, and $t_{z}$ (bottom panels) of a total number of
    $3536$ magnetograms for all the analyzed ARs.}
  \label{hist}
\end{figure*}

\Fig~\ref{hist} shows the histograms of normalized forces and torques
for all the $3536$ magnetograms with both the average and
median values denoted. The averages and standard deviations for the forces are respectively,
$f_x = 0.15 \pm 0.10$, $f_y = 0.13 \pm 0.08$, and $f_z = 0.13\pm 0.08$. Thus
all of them are close to $0.1$, at which the field can be regarded as force-free~\citep{Metcalf1995}.
The median values of the forces are systematically smaller by a little bit than their average values.
The averages and standard deviations for the normalized torques are respectively,
$t_x = 0.05 \pm 0.03$, $t_y = 0.11 \pm 0.07$, and $t_z = 0.12\pm 0.09$.
The torques show a slight dependence on the direction; $t_x$ is approximately a half
of $t_y$ and $t_z$. In any case, on average, both the relative forces and
torques of the magnetograms are close to $0.1$, thus they can be
regarded as being close to force-free. We note that with such a much larger set of emerging AR
samples than previous statistical studies using the snapshots of developed ARs~\citep{moon_forcefreeness_2002, tiwari_force-free_2012},
our results confirm that the photospheric field is actually not far from the force-free state, even in the dynamically emerging phase.

\begin{figure*}[htbp]
  \centering
  \includegraphics[width=0.9\textwidth]{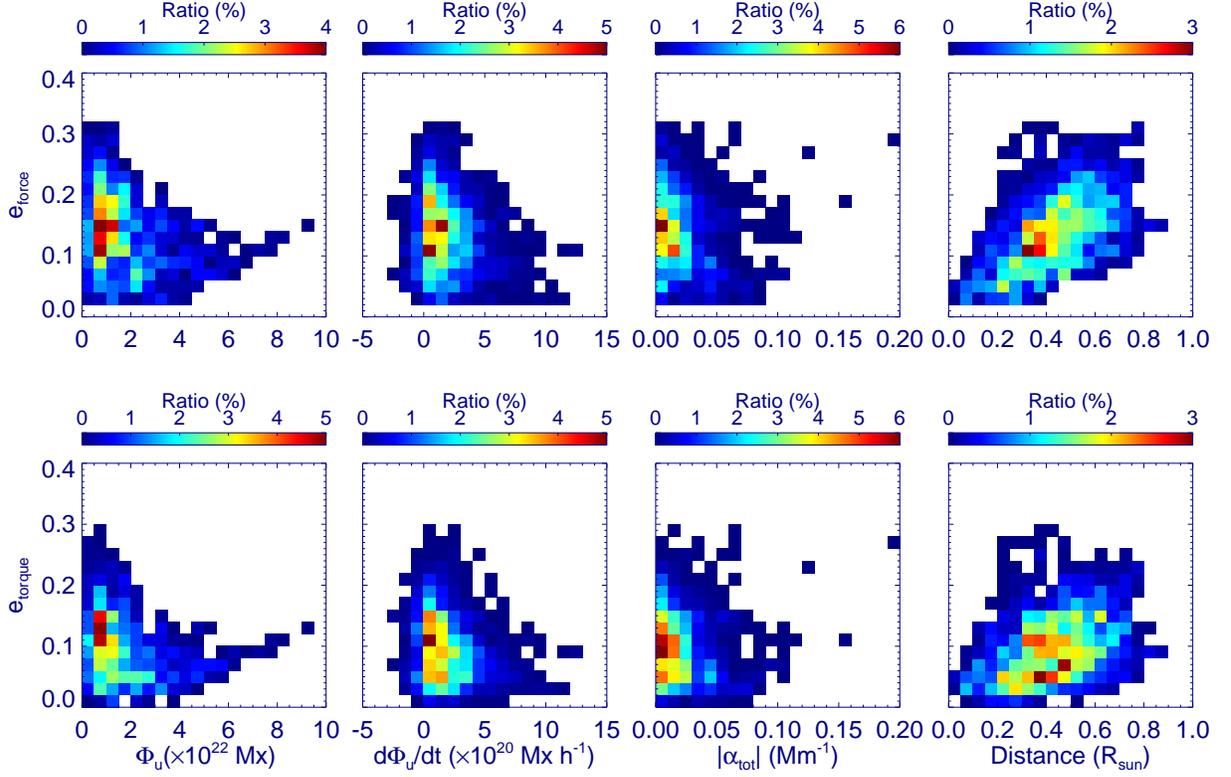}
  \caption{Distributions of magnetic force and magnetic torque with
    different parameters. The top panels are for magnetic force, and
    the bottom ones are for magnetic torque. The horizontal axes from the left to right columns represent, respectively, the total unsigned flux $\Phi_u$, the flux changing rate $d \Phi_u/dt$, the mean
    twist parameter $\alpha_{\rm tot}$, and distance of the ARs from the solar disk center (in unit of solar radius $R_{\rm sun}$).
    The colors indicate ratios of events in a specific bin. Here
    the bins for magnetic force and magnetic torque are both 0.02; and
    for $\Phi_u$, $d \Phi_u/dt$, $\alpha_{\rm tot}$ and distance, they are
    $0.5\times10^{22}$~Mx, $1\times10^{20}$~Mx~h$^{-1}$,
    $0.01$~Mm$^{-1}$ and $0.02 R_{\rm sun}$, respectively.}
  \label{2d}
\end{figure*}

To explore whether there are correlations between the forces (and torques) and key parameters
during the flux emergence of ARs, such as the total unsigned magnetic flux $\Phi_{\rm u}$, the
flux changing rate $d \Phi_{\rm u}/d t$ and the non-potentiality, which is quantified by the
average twist parameter $\alpha_{\rm tot}$~\citep{Bobra2014} defined by
\begin{equation}\label{alpha}
  \alpha_{\rm tot} = \frac{\int J_z B_z dx dy}{\int B_z^2 dx dy}.
\end{equation}
The simple intuition is that, the faster the flux emerges, the larger the force should be;
the more non-potential or twisted the emerging field is, the larger the force and torque should be.
In \Fig~\ref{2d}, we plot the two-dimensional
histogram showing the event frequency distributions in the two-parameter spaces defined by the force (and torque) with the three parameters $\Phi_{\rm u}$, $d \Phi_{\rm u}/d t$, and $\alpha_{\rm tot}$, respectively.
For simplicity, as we only care about the magnitudes, we use average of the three components of force and torque by $e_{\rm force}=(f_x+f_y+f_z)/3$ and $e_{\rm torque}=(t_x+t_y+t_z)/3$.
As can be seen, no systematic correlation is found
between the forces (and torques) with either the total magnetic flux of the
ARs, the emergence rates or the non-potentiality.

We further analyzed whether the force and torque are
influenced by the observed locations of the ARs, since the further away from the solar disk center the target AR is, the larger the measurement errors would be. Indeed, as shown in the last column of \Fig~\ref{2d},  an approximately linear correlation is seen in the force (and torque) with the distance
of the observed AR from the disk center. It shows that the closer to
the disk center, the smaller the force (and torque) is. Thus, the increase of the forces (and torques) is most likely due to the errors in the observations, and with a
better quality of observed data, presumably the force and torque will be smaller, meaning
that the emerging field is actually even closer to force-free than than the results shown here.

\section{Discussion}
\label{sec:dis}
In this Letter, we have, for the first time, systematically surveyed
the normalized magnitudes of the global Lorentz forces and torques in
solar flux-emergence ARs observed by \SDO/HMI. It is found that even during
the flux-emergence phase, i.e., the formation of ARs, the
magnetic field in the photosphere is generally close to force-free
state, since the relative measurements of the Lorentz force and torque
are mostly on the order of $0.1$. There seems to be no correlation between the magnitude of normalized force and torque with the total unsigned flux content of the
emerging ARs, the emergence rate, or the non-potentiality of the field.
The only systematic correlation of the force (and torque) is found to be with the observed location of
the ARs on the solar disk, that is, the closer to the solar disk
center, the closer to force-free the field is. Therefore, this suggests that
the actual magnetic field is even more force-free than what the
statistic results show here. It can explain why many nonlinear force-free coronal-field extrapolations based
directly on the photospheric vector magneotgrams generally agrees with each other to some extent~\citep[e.g.,][]{Wiegelmann2004,Valori2007,Wiegelmann2012solar, Jiang2013NLFFF,Inoue2014}, although there are still some mismatches between the models, and the mismatches can be further alleviated through a preprocessing of the vector magnetograms, in which the photospheric fields are able to be rendered highly force-free by adjusting the observed data within the noise level~\citep{Wiegelmann2006b, Jiang2014Prep, duan_preprocessing_2018}.

\begin{figure*}[htbp]
  \centering
  \includegraphics[width=0.9\textwidth]{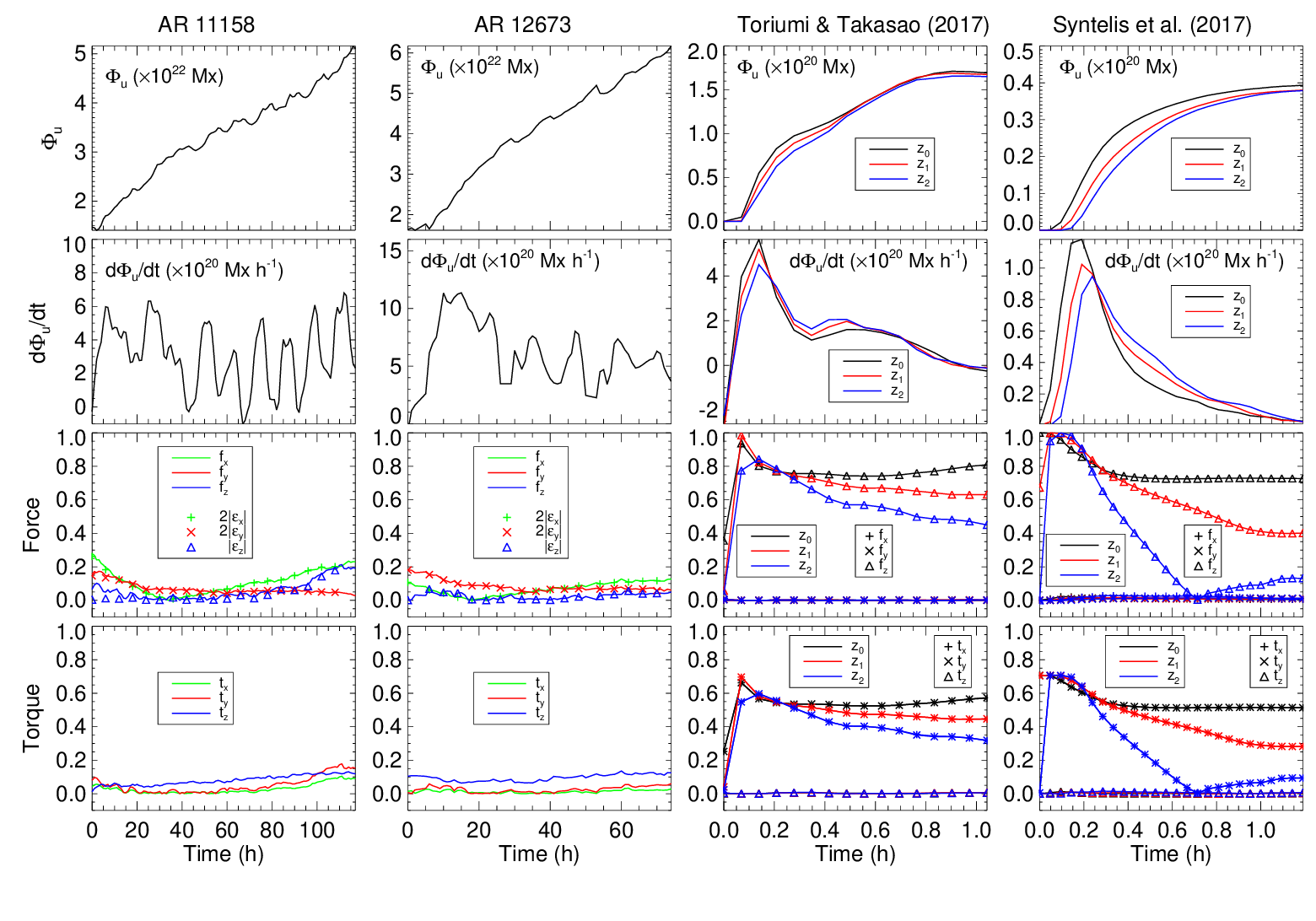}
  \caption{Comparison of different parameters from two observed flux-emerging ARs
    and numerical simulations. From top
    to bottom are, respectively, evolution of the total unsigned flux, flux
    emergence rate, normalized Lorentz force and torque. From left to
    right are results for AR~11158, AR~12673, simulations of \citet{toriumi_numerical_2017} and \citet{syntelis_recurrent_2017}, respectively. For the simulations we show the results for three different heights, which are $(z_0, z_1, z_2) =(0.14, 0.28, 0.42)$~Mm in \citet{toriumi_numerical_2017}'s simulation, and $(z_0, z_1, z_2) =(0.12, 0.28, 0.43)$~Mm in \citet{syntelis_recurrent_2017}'s simulation.
    The horizontal axes show the
    time for different events. Note that in the plots of normalized forces for the two ARs, the results from the \texttt{cgem.Lorentz} data set, i.e., $\epsilon_x$,  $\epsilon_y$, and $\epsilon_z$, defined in \citet{SunX2019} are also shown for a double-checking of our calculations.}
  \label{compare}
\end{figure*}

Our results show that during the emergence phase, idealized flux emergence simulations in which a twisted flux tube is artificially made buoyant to rise through the convection zone, show significantly larger forces  (in comparison to the total magnetic pressure force) at the photosphere than the ones measured by our observations. It is after the flux injection phase, when the photospheric flux saturates, that the models tend to settle to more realistic photospheric force and torque ratios. This is suggestive of a discrepancy between simulations and observations.
Very recently, using the the flux emergence simulation of \citet{toriumi_numerical_2017} as a ground-truth data set (the same one we have analyzed in this Letter as shown in \Fig~\ref{compare}), \citet{Toriumi2020} performed a joint comparison of different data-driven coronal field evolution models that using the photospheric magnetograms produced in the simulated  flux emergence as input to their bottom boundaries~\citep{Cheung2012, Jiang2016NC, Hayashi2019, GuoY2019}. It was found that, although all the data-driven models reproduced a flux rope structure, the quantitative discrepancies are large, which is attributed mainly to the highly non-force-free input photospheric field and to the treatment of background atmosphere. Especially, for a data-driven MHD model~\citep{Jiang2016NC} that \textbf{used} typical settings of the tenuous atmosphere in the corona (i.e., very low plasma $\beta$ and high Alfv{\'e}n speed), the reproduced magnetic flux rope is significantly larger in size and stronger in field-line twisting than those in the original simulation as well as other data-driven MHD models that use typically dense plasma near the lower boundary~\citep[e.g.,][]{GuoY2019, Hayashi2019}. This discrepancy clearly arises from too strong Lorentz force of the simulated photospheric field, which cannot be balanced by the plasma in \citet{Jiang2016NC}'s model, and thus the flux rope can rise strongly upward and be freely twisted by the Lorentz force and torque, which leads to the strong magnetic energy and helicity in the corona. Thus, a further test of the data-driven models with more realistic and thus more force-free photospheric magnetic field \textbf{needs} to be done in future.

By comparing the flux emergence rates from observations with those from typical MHD simulations, \citet{norton_magnetic_2017} have found that the observed emergence rates are smaller than those in simulations, which indicates a slower rise of the flux in the interior than what is captured in simulations. That finding is consistent with ours, since with a slower emergence rate, the emerging field at the photosphere have more time to relax, and consequently, be more force-free. In other words, the emergence at the photosphere might actually proceed in a quasi-static way rather than the dynamic one as shown in simulations. There are several aspects that can be adjusted in simulations to make the field emerge slower and thus, potentially, closer to force-free (and torque-free) in the photosphere. The first one is the depth where the initial flux tube is placed; both the simulations we used here have a flux tube placed near the solar surface, while with a flux tube placed much deeper in the interior, it can expand and stretch with more time to relax during its rising and should become more force-free than initially~\citep[e.g.,][]{Toriumi2011, Syntelis2019S}. The second one is the twist degree of the initial flux tube; stronger twist can of course create increased torque during its emerging in the photosphere~\citep{Sturrock2015, Sturrock2016}. The third one might be attributed to geometry of the emerging tube and how it couples with twist and the size of the flux tube, which has been discussed in~\citet{Syntelis2019S}. Finally, to include realistic turbulent convection in the modeling can further yield much relaxed emergence process~\citep{Toriumi2019S}. In summary, our study shows that the photosphere field is very close to force-free during the emergence process, and this fact should be taken into consideration in future development of MHD simulations as well as the theories of flux emergence.


\acknowledgments

This work is supported by the startup funding (74110-18841214) from Sun
Yat-sen University.  C.J. acknowledges support by National Natural
Science Foundation of China (41822404, 41731067, 41574170, 41531073).
 S.T. was supported by JSPS KAKENHI Grant Numbers JP15H05814 (PI: K. Ichimoto) and JP18H05234 (PI: Y. Katsukawa), and by the NINS program for cross-disciplinary study (Grant Numbers 01321802 and 01311904) on Turbulence, Transport, and Heating Dynamics in Laboratory and Solar/Astrophysical Plasmas: "SoLaBo-X". P.S. acknowledges support by the ERC synergy grant ``The Whole Sun''. Data from observations are courtesy of NASA \SDO/HMI science teams.
 We are grateful to the anonymous referee for his/her comments and suggestions in improving the paper.


\begin{thebibliography}{}
\expandafter\ifx\csname natexlab\endcsname\relax\def\natexlab#1{#1}\fi
\providecommand{\url}[1]{\href{#1}{#1}}
\providecommand{\dodoi}[1]{doi:~\href{http://doi.org/#1}{\nolinkurl{#1}}}
\providecommand{\doeprint}[1]{\href{http://ascl.net/#1}{\nolinkurl{http://ascl.net/#1}}}
\providecommand{\doarXiv}[1]{\href{https://arxiv.org/abs/#1}{\nolinkurl{https://arxiv.org/abs/#1}}}

\bibitem[{Acheson(1979)}]{Acheson1979}
Acheson, D.~J. 1979, Solar Physics, 62, 23, \dodoi{10.1007/BF00150129}

\bibitem[{Aly(1984)}]{aly_properties_1984}
Aly, J.~J. 1984, The Astrophysical Journal, 283, 349, \dodoi{10.1086/162313}

\bibitem[{{Aly}(1989)}]{Aly1989}
{Aly}, J.~J. 1989, \solphys, 120, 19, \dodoi{10.1007/BF00148533}

\bibitem[{{Archontis}(2008)}]{Archontis2008}
{Archontis}, V. 2008, Journal of Geophysical Research (Space Physics), 113, 3,
  \dodoi{10.1029/2007JA012422}

\bibitem[{Archontis {et~al.}(2004)Archontis, Moreno-Insertis, Galsgaard, Hood,
  \& O'Shea}]{archontis_emergence_2004}
Archontis, V., Moreno-Insertis, F., Galsgaard, K., Hood, A., \& O'Shea, E.
  2004, Astronomy \& Astrophysics, 426, 1047,
  \dodoi{10.1051/0004-6361:20035934}

\bibitem[{Archontis \& Syntelis(2019)}]{archontis_emergence_2019}
Archontis, V., \& Syntelis, P. 2019, Philosophical Transactions of the Royal
  Society A: Mathematical, Physical and Engineering Sciences, 377, 20180387,
  \dodoi{10.1098/rsta.2018.0387}

\bibitem[{{Bobra} {et~al.}(2014){Bobra}, {Sun}, {Hoeksema}, {Turmon}, {Liu},
  {Hayashi}, {Barnes}, \& {Leka}}]{Bobra2014}
{Bobra}, M.~G., {Sun}, X., {Hoeksema}, J.~T., {et~al.} 2014, \solphys, 289,
  3549, \dodoi{10.1007/s11207-014-0529-3}

\bibitem[{Cheung \& DeRosa(2012)}]{Cheung2012}
Cheung, M. C.~M., \& DeRosa, M.~L. 2012, The Astrophysical Journal, 757, 147,
  \dodoi{10.1088/0004-637X/757/2/147}

\bibitem[{Cheung \& Isobe(2014)}]{Cheung2014}
Cheung, M. C.~M., \& Isobe, H. 2014, Living Reviews in Solar Physics, 11,
  \dodoi{10.12942/lrsp-2014-3}

\bibitem[{Cho {et~al.}(2017)Cho, Cho, Bong, Moon, Nakariakov, Park, Baek, Choi,
  Kim, \& Lee}]{cho_determination_2017}
Cho, I.-H., Cho, K.-S., Bong, S.-C., {et~al.} 2017, The Astrophysical Journal,
  837, L11, \dodoi{10.3847/2041-8213/aa611b}

\bibitem[{Duan \& Zhang(2018)}]{duan_preprocessing_2018}
Duan, A.-Y., \& Zhang, H. 2018, Research in Astronomy and Astrophysics, 18,
  085, \dodoi{10.1088/1674-4527/18/7/85}

\bibitem[{{Fan}(2009)}]{Fan2009}
{Fan}, Y. 2009, \apj, 697, 1529, \dodoi{10.1088/0004-637X/697/2/1529}

\bibitem[{{Fisher} {et~al.}(2012){Fisher}, {Bercik}, {Welsch}, \&
  {Hudson}}]{Fisher2012}
{Fisher}, G.~H., {Bercik}, D.~J., {Welsch}, B.~T., \& {Hudson}, H.~S. 2012,
  \solphys, 277, 59, \dodoi{10.1007/s11207-011-9907-2}

\bibitem[{{Gary}(2001)}]{Gary2001}
{Gary}, G.~A. 2001, \solphys, 203, 71

\bibitem[{Guo {et~al.}(2019)Guo, Xia, Keppens, Ding, \& Chen}]{GuoY2019}
Guo, Y., Xia, C., Keppens, R., Ding, M.~D., \& Chen, P.~F. 2019, The
  Astrophysical Journal, 870, L21, \dodoi{10.3847/2041-8213/aafabf}

\bibitem[{Hayashi {et~al.}(2019)Hayashi, Feng, Xiong, \& Jiang}]{Hayashi2019}
Hayashi, K., Feng, X., Xiong, M., \& Jiang, C. 2019, The Astrophysical Journal,
  871, L28, \dodoi{10.3847/2041-8213/aaffcf}

\bibitem[{{Hoeksema} {et~al.}(2014){Hoeksema}, {Liu}, {Hayashi}, {Sun},
  {Schou}, {Couvidat}, {Norton}, {Bobra}, {Centeno}, {Leka}, {Barnes}, \&
  {Turmon}}]{Hoeksema2014}
{Hoeksema}, J.~T., {Liu}, Y., {Hayashi}, K., {et~al.} 2014, \solphys, 289,
  3483, \dodoi{10.1007/s11207-014-0516-8}

\bibitem[{Inoue {et~al.}(2014)Inoue, Hayashi, Magara, Choe, \&
  Park}]{Inoue2014}
Inoue, S., Hayashi, K., Magara, T., Choe, G.~S., \& Park, Y.~D. 2014, The
  Astrophysical Journal, 788, 182.
\newblock \url{http://stacks.iop.org/0004-637X/788/i=2/a=182}

\bibitem[{{Jiang} \& {Feng}(2013)}]{Jiang2013NLFFF}
{Jiang}, C., \& {Feng}, X. 2013, \apj, 769, 144,
  \dodoi{10.1088/0004-637X/769/2/144}

\bibitem[{{Jiang} \& {Feng}(2014)}]{Jiang2014Prep}
---. 2014, \solphys, 289, 63, \dodoi{10.1007/s11207-013-0346-0}

\bibitem[{Jiang \& Zhang(2019)}]{jiang_influence_2019}
Jiang, C.-q., \& Zhang, M. 2019, Chinese Astronomy and Astrophysics, 43, 252,
  \dodoi{10.1016/j.chinastron.2019.04.009}

\bibitem[{{Jiang} {et~al.}(2016){Jiang}, {Wu}, {Feng}, \& {Hu}}]{Jiang2016NC}
{Jiang}, C.~W., {Wu}, S.~T., {Feng}, X.~S., \& {Hu}, Q. 2016, Nature Comm., 7,
  11522, \dodoi{10.1038/ncomms11522}

\bibitem[{Liu \& Hao(2015)}]{liu_research_2015}
Liu, S., \& Hao, J. 2015, Advances in Space Research, 55, 1563,
  \dodoi{10.1016/j.asr.2015.01.010}

\bibitem[{Liu {et~al.}(2013)Liu, Su, Zhang, Deng, Gao, Yang, \&
  Mao}]{liu_statistical_2013}
Liu, S., Su, J.~T., Zhang, H.~Q., {et~al.} 2013, Publications of the
  Astronomical Society of Australia, 30, e005, \dodoi{10.1017/pasa.2012.005}

\bibitem[{Low(1985)}]{Low1985}
Low, B.~C. 1985, in Measurements of {Solar} {Vector} {Magnetic} {Fields}, Vol.
  2374 (ed. M. J. Hagyard (Washington, DC: NASA)), 49

\bibitem[{{Metcalf} {et~al.}(1995){Metcalf}, {Jiao}, {McClymont}, {Canfield},
  \& {Uitenbroek}}]{Metcalf1995}
{Metcalf}, T.~R., {Jiao}, L., {McClymont}, A.~N., {Canfield}, R.~C., \&
  {Uitenbroek}, H. 1995, \apj, 439, 474, \dodoi{10.1086/175188}

\bibitem[{{Molodenskii}(1969)}]{Molodenskii1969}
{Molodenskii}, M.~M. 1969, \sovast, 12, 585

\bibitem[{Moon {et~al.}(2002)Moon, Choe, Yun, Park, \&
  Mickey}]{moon_forcefreeness_2002}
Moon, Y.-J., Choe, G.~S., Yun, H.~S., Park, Y.~D., \& Mickey, D.~L. 2002, The
  Astrophysical Journal, 568, 422, \dodoi{10.1086/338891}

\bibitem[{Newcomb(1961)}]{newcomb_convective_1961}
Newcomb, W.~A. 1961, Physics of Fluids, 4, 391, \dodoi{10.1063/1.1706342}

\bibitem[{Norton {et~al.}(2017)Norton, Jones, Linton, \&
  Leake}]{norton_magnetic_2017}
Norton, A.~A., Jones, E.~H., Linton, M.~G., \& Leake, J.~E. 2017, The
  Astrophysical Journal, 842, 3, \dodoi{10.3847/1538-4357/aa7052}

\bibitem[{Otsuji {et~al.}(2011)Otsuji, Kitai, Ichimoto, \&
  Shibata}]{otsuji_statistical_2011}
Otsuji, K., Kitai, R., Ichimoto, K., \& Shibata, K. 2011, Publications of the
  Astronomical Society of Japan, 63, 1047, \dodoi{10.1093/pasj/63.5.1047}

\bibitem[{Parker(1955)}]{Parker1955}
Parker, E.~N. 1955, The Astrophysical Journal, 122, 293, \dodoi{10.1086/146087}

\bibitem[{Shibata {et~al.}(1989)Shibata, Tajima, Matsumoto, Horiuchi, Hanawa,
  Rosner, \& Uchida}]{shibata_nonlinear_1989}
Shibata, K., Tajima, T., Matsumoto, R., {et~al.} 1989, The Astrophysical
  Journal, 338, 471, \dodoi{10.1086/167212}

\bibitem[{Sturrock \& Hood(2016)}]{Sturrock2016}
Sturrock, Z., \& Hood, A.~W. 2016, Astronomy \& Astrophysics, 593, A63,
  \dodoi{10.1051/0004-6361/201628360}

\bibitem[{Sturrock {et~al.}(2015)Sturrock, Hood, Archontis, \&
  McNeill}]{Sturrock2015}
Sturrock, Z., Hood, A.~W., Archontis, V., \& McNeill, C.~M. 2015, Astronomy \&
  Astrophysics, 582, A76, \dodoi{10.1051/0004-6361/201526521}

\bibitem[{{Sun}(2014)}]{SunX2019}
{Sun}, X. 2014, arXiv e-prints, arXiv:1405.7353.
\newblock \doarXiv{1405.7353}

\bibitem[{{Sun} {et~al.}(2012){Sun}, {Hoeksema}, {Liu}, {Wiegelmann},
  {Hayashi}, {Chen}, \& {Thalmann}}]{Sun2012}
{Sun}, X., {Hoeksema}, J.~T., {Liu}, Y., {et~al.} 2012, \apj, 748, 77,
  \dodoi{10.1088/0004-637X/748/2/77}

\bibitem[{{Syntelis} {et~al.}(2019){Syntelis}, {Archontis}, \&
  {Hood}}]{Syntelis2019S}
{Syntelis}, P., {Archontis}, V., \& {Hood}, A. 2019, \apj, 874, 15,
  \dodoi{10.3847/1538-4357/ab0959}

\bibitem[{Syntelis {et~al.}(2017)Syntelis, Archontis, \&
  Tsinganos}]{syntelis_recurrent_2017}
Syntelis, P., Archontis, V., \& Tsinganos, K. 2017, The Astrophysical Journal,
  850, 95, \dodoi{10.3847/1538-4357/aa9612}

\bibitem[{Tiwari(2012)}]{tiwari_force-free_2012}
Tiwari, S.~K. 2012, The Astrophysical Journal, 744, 65,
  \dodoi{10.1088/0004-637X/744/1/65}

\bibitem[{{Toriumi} \& {Hotta}(2019)}]{Toriumi2019S}
{Toriumi}, S., \& {Hotta}, H. 2019, \apjl, 886, L21,
  \dodoi{10.3847/2041-8213/ab55e7}

\bibitem[{Toriumi \& Takasao(2017)}]{toriumi_numerical_2017}
Toriumi, S., \& Takasao, S. 2017, The Astrophysical Journal, 850, 39,
  \dodoi{10.3847/1538-4357/aa95c2}

\bibitem[{{Toriumi} {et~al.}(2020){Toriumi}, {Takasao}, {Cheung}, {Jiang},
  {Guo}, {Hayashi}, \& {Inoue}}]{Toriumi2020}
{Toriumi}, S., {Takasao}, S., {Cheung}, M. C.~M., {et~al.} 2020, \apj, 890,
  103, \dodoi{10.3847/1538-4357/ab6b1f}

\bibitem[{Toriumi \& Yokoyama(2010)}]{Toriumi2010}
Toriumi, S., \& Yokoyama, T. 2010, The Astrophysical Journal, 714, 505,
  \dodoi{10.1088/0004-637X/714/1/505}

\bibitem[{Toriumi \& Yokoyama(2011)}]{Toriumi2011}
---. 2011, The Astrophysical Journal, 735, 126,
  \dodoi{10.1088/0004-637X/735/2/126}

\bibitem[{{Valori} {et~al.}(2007){Valori}, {Kliem}, \& {Fuhrmann}}]{Valori2007}
{Valori}, G., {Kliem}, B., \& {Fuhrmann}, M. 2007, \solphys, 245, 263,
  \dodoi{10.1007/s11207-007-9046-y}

\bibitem[{{Wiegelmann}(2004)}]{Wiegelmann2004}
{Wiegelmann}, T. 2004, \solphys, 219, 87,
  \dodoi{10.1023/B:SOLA.0000021799.39465.36}

\bibitem[{{Wiegelmann} {et~al.}(2006){Wiegelmann}, {Inhester}, \&
  {Sakurai}}]{Wiegelmann2006b}
{Wiegelmann}, T., {Inhester}, B., \& {Sakurai}, T. 2006, \solphys, 233, 215,
  \dodoi{10.1007/s11207-006-2092-z}

\bibitem[{{Wiegelmann} \& {Sakurai}(2012)}]{Wiegelmann2012solar}
{Wiegelmann}, T., \& {Sakurai}, T. 2012, Living Reviews in Solar Physics, 9, 5.
\newblock \doarXiv{1208.4693}

\bibitem[{{Yan} {et~al.}(2018){Yan}, {Wang}, {Pan}, {Kong}, {Xue}, {Yang},
  {Li}, \& {Feng}}]{YanX2018}
{Yan}, X.~L., {Wang}, J.~C., {Pan}, G.~M., {et~al.} 2018, \apj, 856, 79,
  \dodoi{10.3847/1538-4357/aab153}

\bibitem[{{Yang} {et~al.}(2017){Yang}, {Zhang}, {Zhu}, \& {Song}}]{YangS2017}
{Yang}, S., {Zhang}, J., {Zhu}, X., \& {Song}, Q. 2017, \apjl, 849, L21,
  \dodoi{10.3847/2041-8213/aa9476}

\end{thebibliography}

\clearpage

\begin{longtable*}{cccccccc}
  \caption{NOAA numbers, durations, and locations of all the studied
    flux emerging ARs. The longitude is Carrington longitude of the
    AR's center (i.e., the center of the SHARP patch) with respect to
    the disk center, in unit of degree. The latitude is Carrington
    latitude of the AR's center, in unit of degree.}\\
 \hline
  \hline
  No. & NOAA AR   & Start Time & Longitude & Latitude & End Time & Longitude  & Latitude \\
  \hline
   1 & AR11072 & 2010-05-21T08:00 & -25.29 & -13.63 & 2010-05-26T13:00 &  44.81 & -14.24 \\
    2 & AR11076 & 2010-05-31T20:00 &  -6.78 & -18.84 & 2010-06-04T16:00 &  44.47 & -19.31 \\
    3 & AR11117 & 2010-10-26T04:00 &   4.30 &  18.27 & 2010-10-29T05:00 &  44.66 &  18.56 \\
    4 & AR11130 & 2010-11-28T08:00 & -10.53 &  11.88 & 2010-12-02T10:00 &  44.58 &  12.39 \\
    5 & AR11141 & 2010-12-30T22:00 &  -3.05 &  37.20 & 2011-01-02T02:00 &  24.93 &  37.46 \\
    6 & AR11158 & 2011-02-12T14:00 & -20.21 & -14.39 & 2011-02-17T11:00 &  44.85 & -14.17 \\
    7 & AR11327 & 2011-10-20T14:00 & -18.73 & -26.34 & 2011-10-24T01:00 &  27.38 & -26.06 \\
    8 & AR11416 & 2012-02-08T14:00 & -44.02 & -11.58 & 2012-02-15T05:00 &  44.72 & -11.22 \\
    9 & AR11422 & 2012-02-19T12:00 &  -9.89 &  22.59 & 2012-02-22T23:00 &  36.62 &  22.71 \\
    10 & AR11431 & 2012-03-04T12:00 &  16.26 & -20.95 & 2012-03-06T16:00 &  44.67 & -20.95 \\
    11 & AR11460 & 2012-04-18T01:00 & -25.59 &  21.47 & 2012-04-23T00:00 &  41.11 &  21.04 \\
    12 & AR11551 & 2012-08-20T04:00 & -11.20 &   5.26 & 2012-08-23T21:00 &  38.92 &   5.14 \\
    13 & AR11561 & 2012-08-30T01:00 & -28.89 & -19.08 & 2012-08-30T20:00 & -18.18 & -19.08 \\
    14 & AR11630 & 2012-12-08T12:00 & -23.39 &  19.09 & 2012-12-11T10:00 &  15.65 &  19.47 \\
    15 & AR11640 & 2012-12-30T14:00 & -23.17 &  30.84 & 2013-01-04T18:00 &  44.73 &  31.43 \\
    16 & AR11645 & 2013-01-02T20:00 & -12.58 & -10.24 & 2013-01-05T00:00 &  16.66 &  -9.98 \\
    17 & AR11682 & 2013-02-26T02:00 &  -9.25 & -11.21 & 2013-02-29T00:00 &  29.82 & -11.17 \\
    18 & AR11702 & 2013-03-20T19:00 &  10.00 &  15.09 & 2013-03-23T06:00 &  43.36 &  15.00 \\
    19 & AR11726 & 2013-04-19T06:00 & -14.30 &  18.00 & 2013-04-23T14:00 &  44.80 &  17.60 \\
    20 & AR11750 & 2013-05-15T01:00 &   0.95 &  -7.33 & 2013-05-17T16:00 &  36.51 &  -7.63 \\
    21 & AR11764 & 2013-06-02T01:00 &  10.09 &  12.85 & 2013-06-04T00:00 &  36.56 &  12.61 \\
    22 & AR11765 & 2013-06-05T13:00 & -24.40 &  10.51 & 2013-06-10T00:00 &  36.03 &   9.98 \\
    23 & AR11776 & 2013-06-20T05:00 &  12.50 &   9.24 & 2013-06-22T14:00 &  44.62 &   8.98 \\
    24 & AR11781 & 2013-06-27T21:00 & -11.35 &  19.02 & 2013-07-01T00:00 &  30.25 &  18.66 \\
    25 & AR11784 & 2013-07-03T13:00 & -11.31 & -17.98 & 2013-07-05T00:00 &   8.33 & -18.14 \\
    26 & AR11807 & 2013-07-28T11:00 &  -3.17 &  23.35 & 2013-07-30T12:00 &  23.58 &  23.20 \\
    27 & AR11813 & 2013-08-07T12:00 &  -0.32 & -19.55 & 2013-08-10T00:00 &  33.42 & -19.71 \\
    28 & AR11824 & 2013-08-17T11:00 &   4.51 & -20.47 & 2013-08-19T17:00 &  34.87 & -20.56 \\
    29 & AR11843 & 2013-09-17T08:00 & -15.24 &  -6.14 & 2013-09-19T05:00 &  10.28 &  -6.12 \\
    30 & AR11855 & 2013-09-30T15:00 & -20.03 & -20.24 & 2013-10-04T00:00 &  25.52 & -20.09 \\
    31 & AR11922 & 2013-12-10T01:00 &   5.96 &   9.86 & 2013-12-12T22:00 &  44.92 &  10.22 \\
    32 & AR11946 & 2014-01-05T16:00 & -25.21 &  12.20 & 2014-01-10T00:00 &  33.55 &  12.69 \\
    33 & AR12003 & 2014-03-10T15:00 &   4.37 &  15.15 & 2014-03-13T13:00 &  44.00 &  15.12 \\
    34 & AR12089 & 2014-06-13T15:00 &   6.27 &  17.22 & 2014-06-16T12:00 &  44.81 &  16.87 \\
    35 & AR12119 & 2014-07-18T10:00 & -23.26 & -25.92 & 2014-07-23T00:00 &  37.78 & -26.33 \\
    36 & AR12219 & 2014-11-25T13:00 & -13.48 &   2.72 & 2014-11-29T20:00 &  44.89 &   3.24 \\
    37 & AR12234 & 2014-12-12T06:00 &   1.81 &   4.56 & 2014-12-15T00:00 &  39.22 &   4.90 \\
    38 & AR12257 & 2015-01-09T03:00 &  16.99 &   9.96 & 2015-01-10T20:00 &  40.20 &  10.13 \\
    39 & AR12273 & 2015-01-26T12:00 &  -8.98 &   2.86 & 2015-01-29T14:00 &  32.98 &   3.10 \\
    40 & AR12422 & 2015-09-24T19:00 & -26.50 & -26.83 & 2015-09-30T03:00 &  44.73 & -26.63 \\
    41 & AR12530 & 2016-04-11T00:00 &  -2.54 &  21.03 & 2016-04-13T00:00 &  24.38 &  20.88 \\
    42 & AR12543 & 2016-05-09T06:00 & -11.00 &  -2.04 & 2016-05-12T21:00 &  38.28 &  -2.42 \\
    43 & AR12571 & 2016-08-05T18:00 & -12.30 &   7.86 & 2016-08-09T23:00 &  44.47 &   7.59 \\
    44 & AR12581 & 2016-08-29T22:00 &  19.07 &   5.11 & 2016-08-31T06:00 &  37.08 &   5.09 \\
    45 & AR12635 & 2017-02-08T17:00 & -31.45 &  19.54 & 2017-02-13T09:00 &  31.55 &  19.79 \\
    46 & AR12673 & 2017-09-03T04:00 &  -8.37 & -16.47 & 2017-09-06T06:00 &  33.44 & -16.49 \\
    47 & AR12675 & 2017-08-30T23:00 &   8.50 & -13.21 & 2017-09-01T11:00 &  28.88 & -13.21 \\
    48 & AR12715 & 2018-06-19T11:00 & -40.22 &   6.69 & 2018-06-20T22:00 & -20.43 &   6.53 \\
    49 & AR12720 & 2018-08-23T21:00 &  10.78 &   0.83 & 2018-08-25T16:00 &  35.10 &   0.79 \\
    50 & AR12723 & 2018-09-29T15:00 &  -4.16 & -16.11 & 2018-10-02T08:00 &  32.55 & -15.98 \\
    51 & AR12733 & 2019-01-24T14:00 &   3.12 &  11.06 & 2019-01-26T21:00 &  34.26 &  11.26 \\
    \hline
    \label{tab1}
\end{longtable*}

\end{document}